\documentclass{aa}
\usepackage{graphicx}
\usepackage{xcolor}
\usepackage{soul}
\usepackage{txfonts}
\graphicspath{ {figures/} }

\def\Gaia{\textit{Gaia}\xspace}
\newcommand{\wjk}{\ensuremath{W_{\rm J,K_s}}\xspace}
\newcommand{\wbr}{\ensuremath{W_{\rm BP,RP}}\xspace}
\newcommand{\wkr}{\ensuremath{W_{\rm K_s,RP}}\xspace}
\newcommand{\bpminrp}{\ensuremath{G_\mathrm{BP}-G_\mathrm{RP}}\xspace}

\begin{document} 

\title{The evolutionary state of the red giant star L$_2$ Puppis\thanks{Based on observations collected at the European Organisation for Astronomical Research in the Southern Hemisphere under ESO programmes 072.D-0235(A) and 094.A-9029(E).}}

\author{
S.~Uttenthaler\inst{\ref{inst_iap}}
}

\institute{
Institute of Applied Physics, TU Wien, Wiedner Hauptstr.\ 8-10, 1040 Vienna, Austria; {\tt stefan.uttenthaler@gmail.com}\label{inst_iap}
}

\date{Received September 08, 2024; accepted November 17, 2024}

\abstract
{L$_2$ Puppis (L$_2$~Pup) is a nearby red giant star and an important object in late-type star research because it has a dust disc and potentially a companion.}
{L$_2$~Pup is often called the second-closest Asymptotic Giant Branch (AGB) star to the sun, second only to R~Doradus. However, whether the star is indeed on the AGB or the Red Giant Branch (RGB) is questionable. We review its evolutionary state.}
{We analyse high-resolution optical archive spectra to search for absorption lines of the third dredge-up indicator technetium (Tc) in L$_2$~Pup. We also compare the star to a sample of well-known AGB stars in terms of luminosity and pulsation properties and place it in a \Gaia-2MASS diagram.}
{L$_2$~Pup is found to be Tc-poor. Thus, it is not undergoing third dredge-up events. The star is fainter than the RGB tip and fainter than all Tc-rich stars in the comparison sample. L$_2$~Pup pulsates in the fundamental mode, similar to Mira variables, but its pulsation properties do not allow us to distinguish between the RGB and AGB stages.}
{In conclusion, L$_2$~Pup could be an RGB or early AGB star, but it is more likely to be an RGB than an AGB star. Our findings are important for a better understanding of the L$_2$~Pup system and its past and future evolution.}

\maketitle

\section{Introduction}\label{sec:Intro}

L$_2$~Puppis (L$_2$~Pup) is an important object in the research of late stages of stellar evolution. There are several reasons for its importance: i) It is nearby \citep[$d\approx56$\,pc,][]{2023A&A...674A...1G}; ii) it has a dust disc viewed almost edge-on \citep{2014A&A...564A..88K}; iii) it likely has a stellar \citep{2015A&A...578A..77K} or sub-stellar \citep{2016A&A...596A..92K} companion at 2\,AU projected separation; and iv) it is thought to be in an early stage in the formation of a bipolar planetary nebula \citep{2015A&A...578A..77K}. Consequently, at the time of writing (August 2024), the Simbad astronomical database\footnote{\url{https://simbad.u-strasbg.fr/simbad/}} counts 223 publications about L$_2$~Pup, 77 of which have been published since 2010 alone. In NASA ADS\footnote{\url{https://ui.adsabs.harvard.edu/}}, the numbers are 238 and 87, respectively. Simbad lists L$_2$~Pup as an Asymptotic Giant Branch (AGB) star, the Wikipedia\footnote{\url{https://en.wikipedia.org/wiki/L2_Puppis}} says that it is "most likely" an AGB star, and so do many publications \citep[e.g.,][]{2017A&A...601A...5H,2022MNRAS.510.2363H,2024MNRAS.532..734V}. In particular, \citet[][their Sect.~4.1]{2014A&A...564A..88K} discuss the evolutionary state of L$_2$~Pup and conclude that it is at the beginning of the AGB phase.

Here, we would like to tackle the question of whether L$_2$~Pup is a first-ascent red giant branch (RGB) or a second-ascent AGB star and inspect the associated evidence. This work is motivated by the fact that L$_2$~Pup would be a very low-luminosity technetium-rich outlier if the available literature information were correct. The present paper is, therefore, mainly based on a new analysis of high-resolution optical spectra for the presence of technetium (Tc) in its atmosphere and a comparison to a sample of known AGB stars. Constraints on L$_2$~Pup's evolutionary state are essential to understand better the evolution of the whole system, which consists of the primary, the companion, and the circumstellar disc.

In globular clusters, we can distinguish effectively RGB from AGB stars in HR diagrams using only colours and magnitudes \citep{2013Natur.498..198C}, but the differences are too subtle to be applicable to field stars with a range of metallicities, masses, and other stellar parameters. Therefore, we need other criteria to distinguish the two giant phases. One criterion is the luminosity of stars because AGB stars can become much brighter than the maximum reached on the RGB. This RGB tip luminosity reached before He-core burning (horizontal branch or red clump phase) replaces H-shell burning is well-explained theoretically and takes on a value of $L\approx3000\,L_{\sun}$ for solar metallicity \citep{2008A&A...484..815B}. It depends little on parameters such as the initial mass or the metallicity and is thus routinely used as a standard candle for extra-galactic distance determination \citep{2000ApJS..128..431F}. Thus, we can be sure that stars brighter than the RGB tip are in the AGB phase; below this limit, the star could be on the RGB or (early) AGB.


On the AGB, the star has a double-shell structure consisting of a H-burning shell and a He-burning shell. On the early AGB, both shells burn relatively smoothly, with the H-burning shell increasing in luminosity with time and the He-burning shell slightly decreasing in luminosity as the He-shell becomes thinner. Eventually, thermal instabilities lead to thermonuclear runaway He-shell flash events, also called thermal pulses (TP), that quasi-periodically interrupt the H-burning shell every $10^4$ years or so, depending on mass. This marks the beginning of the TP-AGB phase. If a TP is strong enough, it can be followed by a deep mixing event called the third dredge-up (3DUP) that mixes products of nuclear processing to the surface \citep{2005ARA&A..43..435H}. Most notably, these products are $^{12}$C and elements synthesised by the slow neutron-capture process ($s$-process). One of the $s$-process elements is technetium (Tc, $Z=43$), which has only radioactively unstable isotopes. Its longest-lived isotope produced by the $s$-process, $^{99}$Tc, has a half-life of $\sim2\times10^5$\,yrs, which is short compared to the $\sim10^6$\,yrs a star spends on the upper AGB \citep[e.g.][]{2020MNRAS.498.3283P}. It thus serves as an indicator of recent or ongoing 3DUP in an AGB star \citep{1952ApJ...116...21M,1987AJ.....94..981L} or to discriminate between intrinsic (i.e., through internal nucleosynthesis and mixing) and extrinsic (i.e., through external mass transfer from a former AGB binary companion) $s$-process enrichment \citep{1988ApJ...333..219S,1993A&A...271..463J}. Employing the presence of atmospheric Tc absorption lines in the blue optical range ($\sim4200-4300$\,\AA), one can easily distinguish Tc-rich TP-AGB stars from less evolved phases or stars that are not massive enough to experience 3DUP.

Finally, large-amplitude, long-period pulsation is another important criterion for distinguishing AGB stars from RGB stars. In particular, Mira variables pulsating in the radial fundamental mode \citep[FM,][]{2017ApJ...847..139T} are thought to occupy the AGB. Miras are a sub-class of the long-period variables (LPVs) among luminous red giants and are recognised by periods longer than $\sim100$\,d and amplitudes $\Delta V>2\fm5$ in the visual range \citep{Samus2017}. Semi-regular variables (SRVs) are essentially the extension of the Mira class to smaller amplitude variability of cool giants. They are further subdivided in SRa and SRb types, where the former display persistent and predictable periodicity with amplitudes $\Delta V<2\fm5$, and the latter have poorly defined periodicity. In particular SRa variables can, like the Miras, pulsate in the FM, and the SRb variables are mostly assigned to overtone (OT) modes, where more than one mode can be excited in one star. It is thought that RGB stars preferentially pulsate in one of the OT modes and they can be distinguished from the AGB stars below the RGB tip \citep{2004AcA....54..129S}, but it cannot be excluded that some also pulsate in the FM.


In summary, AGB stars can be unambiguously distinguished by the presence of the 3DUP indicator Tc, luminosities above the RGB tip ($L\gtrsim3000\,L_{\sun}$), and large-amplitude variability. We will investigate below if L$_2$~Pup fulfils these criteria.

The analysis of L$_2$~Pup is complicated by the fact that it entered a long-lasting dimming event in 1995 \citep{2002MNRAS.337...79B}. The dimming event is interpreted to be the result of changes in the line-of-sight opacity of the intervening (disc) material. The Keplerian rotation of the disc might result in time-variable occultations of the central star \citep{2015A&A...578A..77K}. We therefore use near-IR data of L$_2$~Pup from two different sources, obtained before and after the onset of the dimming, respectively. The DIRBE observations in 1989–1990 from \citet{2004ApJS..154..673S}, consisting of 868 individual observations in the $2.2\,\mu{\rm m}$ band, will be adopted to determine L$_2$~Pup's overall luminosity and its $K$-band brightness. According to the relations derived by \citet{2004ApJS..154..673S}, L$_2$~Pup's apparent $2.2\,\mu{\rm m}$ magnitude of $-2\fm376$ can be translated to $K_{\rm S}=-2\fm437$ in the 2MASS system. Thus, the differences between the DIRBE and 2MASS filter systems are small enough to be neglected in this context\footnote{The actually observed post-dimming 2MASS magnitude is $K_{\rm S}=-2\fm308$.}. Note that the DIRBE photometry of the brightest IR sources is ten times more precise than that of 2MASS \citep{2004ApJS..154..673S}. However, the DIRBE photometry is unsuitable to be combined with the post-dimming \Gaia observations. Therefore, we combine the \Gaia photometry with the near-IR photometry of \citet[][their Table~2]{2002MNRAS.337...79B} to be more consistent.

The paper is structured in the following way: In Sect.~\ref{sec:compAGB}, we present the comparison AGB sample. In Sect.~\ref{sec:Tc}, we analyse archive spectra of L$_2$~Pup to revise its Tc classification. The luminosity is analysed and compared in Sect.~\ref{sec:Luminosity}. L$_2$~Pup's pulsation properties are discussed in Sect.~\ref{sec:Pulsation}, and its location in the \Gaia-2MASS diagram in Sect.~\ref{sec:Gaia2MASS}. Finally, conclusions are drawn in Sect.~\ref{sec:Conclusio}.

\section{The comparison sample}\label{sec:compAGB}

We compare L$_2$~Pup to the sample of AGB stars with observations for the 3DUP indicator Tc collected by \citet{2019A&A...622A.120U} in their Table~A.1. A few stars with new Tc observations were added to this sample in the meantime \citep{2024A&A...690A.393U}. We also discuss the main properties of this sample in this paper to put L$_2$~Pup into perspective. De-reddened absolute magnitudes and luminosities for these stars were calculated as follows.

Photometry in 19 different bands was collected that cover the range $0.3 - 90\,\mu{\rm m}$, that is in the $B$, $V$, $I$ (all from AAVSO\footnote{\url{https://www.aavso.org/}}), $J$, $H$, $K_{\rm S}$, $L$, and $M$ bands \citep{1979SAAOC...1...61C,1992A&AS...93..151F,1994A&AS..106..397K,1994MNRAS.267..711W,1995A&AS..113..441K,1996A&A...311..273K,2000MNRAS.319..728W,2004ApJS..154..673S,2006AJ....131.1163S,2006MNRAS.369..751W,2008MNRAS.386..313W,2010ApJS..190..203P,2012A&A...540A..72S}, in the $W1$ to $W4$ bands of the WISE space observatory \citep{2012wise.rept....1C,2019ApJS..240...30S}, IRAS 12, 25, and 60\,$\mu{\rm m}$ \citep{1984ApJ...278L...1N}, and Akari 9, 18, 65, and 90\,$\mu{\rm m}$ bands \citep{2010A&A...514A...1I}. Time-averaged photometry was adopted where available to minimise the effects of stellar variability. Photometry in the $J$-, $H$-, and $K$-bands were converted to the 2MASS system using the relations of \citet{2001AJ....121.2851C}. For the COBE/DIRBE fluxes, the zero-magnitude fluxes as given by \citet{2010ApJS..190..203P} were adopted.

Distances of the stars were calculated by inverting \Gaia DR3 parallaxes \citep{2023A&A...674A...1G}, taking into account the average zero-point offset of quasars of $-21$ microarcseconds as provided by \citet{2021A&A...649A...4L}. Other prescriptions to calculate the zero-point offset are available in the literature \citep[e.g., ][]{2021A&A...654A..20G} but the precise choice is not critical to the results here. For ten sample stars, we adopted the Hipparcos parallaxes of \citet{2007A&A...474..653V} because either the \Gaia parallaxes are missing for the very bright stars or because the Hipparcos parallax is more precise. A limit of $\sigma_{\varpi}/\varpi<0.1$ was applied on the parallax uncertainty to select stars for the comparison here. With this threshold, we obtain a comparison sample of 282 putative AGB stars (a few of them might also be RGB stars).

We also compared the distances obtained from the simple parallax inversion ($d_{\varpi}$) to the distances obtained by the more rigorous treatment of \citet[][$d_{\rm BJ}$]{2021AJ....161..147B}. The average fractional difference $\delta=(d_{\varpi}-d_{\rm BJ})/d_{\varpi}$ is only $+0.005$ and the standard deviation of $\delta$ is 0.017. This translates into an average change in absolute magnitude of $0\fm037$. We consider this as a negligible effect for the purpose of this paper. However, for stars with more uncertain parallaxes ($\sigma_{\varpi}/\varpi>0.1$) or distances greater than $\sim2$\,kpc, the difference in distance can be more substantial.

The photometry was de-reddened using the 3D map of \citet{2017AstL...43..472G}, applying the extinction law of \citet{2016ApJS..224...23X} and adopting the calculated stellar distances. The extinction coefficients for some of the bands that were not determined by \citet{2016ApJS..224...23X} were interpolated from their Fig.~18, whereas no extinction correction was applied to the IRAS 60, Akari 65, and Akari 90\,$\mu{\rm m}$ bands. Since the map of \citet{2017AstL...43..472G} extends only out to 1200\,pc from the sun, extinction was set to this outer boundary for the more distant stars.

Finally, stellar luminosities were calculated by numerical integration of the de-reddened SEDs. On the short-wavelength end, the SED was linearly extrapolated from the $B$-band flux to the point $(\lambda,f_{\lambda})=(0,0)$, and at the long-wavelength end from the available band with the longest effective wavelength (usually IRAS~60 or Akari~90) to the point $(\nu,f_{\nu})=(0,0)$.

The comparison sample is subdivided into the various chemical spectral types, following the well-known evolutionary sequence M--MS--S--C, where the M class is again subdivided into the Tc-poor and Tc-rich groups. The two SC-type stars R~CMi and RZ~Peg are subsumed in the C class. The sample's distribution on the chemical sub-types and their minimum, maximum, and mean luminosities are reported in Table~\ref{tab:lumi}.

\begin{table}
\caption{Luminosities of different groups of (AGB) stars.}\label{tab:lumi}
\centering
\begin{tabular}{lrrrr}
\hline\hline
Group     & $N$ &  $L_{\rm min}$ & $L_{\rm max}$ & $\left<L\right>$        \\
          &     & ($L_{\sun}$)   & ($L_{\sun}$)  & ($L_{\sun}$) \\
\hline
M (no Tc) & 156 &            770 &         34670 & 4290 \\
M (Tc)    &  28 &           2810 &         17270 & 7630 \\
MS (Tc)   &  12 &           2940 &         15580 & 6950 \\
S (Tc)    &  42 &           1800 &         15080 & 5920 \\
C (Tc)    &  44 &           4130 &         27770 & 9300 \\
\hline
\end{tabular}
\end{table}

\section{Technetium and other s-process elements}\label{sec:Tc}

\begin{figure*}[ht!]
\includegraphics[width=\linewidth]{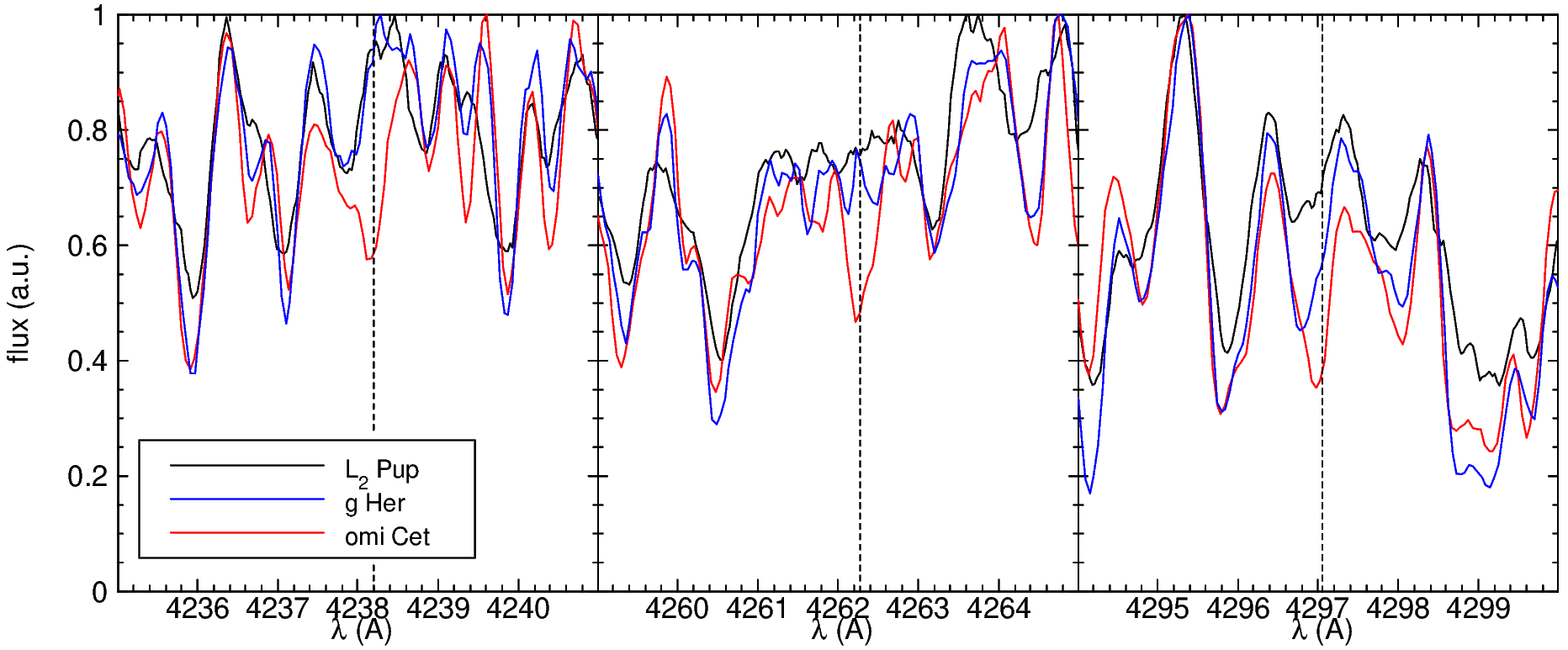}
\caption{Spectra of L$_2$~Pup and the comparison stars $o$~Cet and g~Her around the three Tc resonance transitions at 4238.191 ({\em left panel}), 4262.270 ({\em middle panel}), and 4297.058\,\AA\ ({\em right panel}), indicated by the vertical dashed lines.}
\label{fig:Tclines}
\end{figure*}

\citet{2019A&A...622A.120U} classified L$_2$~Pup as being Tc-rich. This classification was based on \citet{1987AJ.....94..981L}, who classified it as "probably" Tc-rich, and \citet{1999A&A...351..533L}, who classified it as "possibly" Tc-rich. To resolve this ambiguity, we searched the ESO archive for an optical high-resolution spectrum of L$_2$~Pup and found two observations made with the Fibre-fed Extended Range Optical Spectrograph (FEROS, $R=48\,000$) mounted to the 2.2\,m ESO/MPG telescope on La Silla observatory, Chile. The first observation was made in program ID 072.D-0235(A) (PI: C.\ Charbonnel) on 4 January 2004 and had an exposure time of 80\,s. The second one was made in program ID 094.A-9029(E) (PI: R.\ Gredel) on 5 January 2015 with an exposure time of 300\,s. The reduced 2015 spectrum is shown together with spectra of two comparison stars around the three \ion{Tc}{i} resonance lines at 4238.191, 4262.270, and 4297.058\,\AA\ \citep{1968NISTJ..72A.559B} in Fig.~\ref{fig:Tclines}. The comparison stars are $o$~Cet, which is known to be Tc-rich \citep[][and references therein]{1987AJ.....94..981L}, and g~Her, which is known to be Tc-poor \citep{1987AJ.....94..981L,1999A&A...351..533L}. Their spectra have been obtained with the HEROS spectrograph ($R=22\,000$), mounted to the robotic 1.2\,m TIGRE telescope, located near Guanajuato in central Mexico \citep{2014AN....335..787S}. The reduction of the 2004 observation seems to have failed around the Tc line at 4238.191\,\AA, which is why we plot only the 2015 spectrum here. Nevertheless, the spectra are very similar around the two other Tc lines and yield the same result in the Tc classification.

The comparison of L$_2$~Pup's archive spectrum with the other stars in Fig.~\ref{fig:Tclines} shows that it matches very closely the spectrum of the Tc-poor star g~Her. The Tc-rich star $o$~Cet unambiguously has an absorption feature coinciding with the laboratory wavelengths of the 4238 and 4262 \ion{Tc}{i} lines that is lacking in the other two stars (left and middle panel in Fig.~\ref{fig:Tclines}). The \ion{Tc}{i} 4297 is a little less unambiguous because it is blended with other nearby metal lines. Nevertheless, the core of the line blend is shifted to longer wavelengths in $o$~Cet compared to L$_2$~Pup and g~Her. This line shift has often been used to distinguish Tc-rich from Tc-poor stars \citep{1999A&A...351..533L}. Based on these observations, we revise the classification of L$_2$~Pup to be Tc-poor. Therefore, L$_2$~Pup has not undergone 3DUP events in sufficient numbers or efficiency to display Tc lines in its atmosphere. This also makes scenarios for the origin of the disc material that involve TPs unlikely \citep{2016MNRAS.460.4182C}.

Furthermore, Zr is another s-process element and its atmospheric enhancement can be detected via ZrO molecular bands \citep{1979ApJ...234..538A,1979RA......9...39B}. Unlike Tc, Zr has several stable isotopes and thus any enrichment could stem from mass transfer from a former AGB binary companion, which would now be a white dwarf. We searched the FEROS spectra of L$_2$~Pup for ZrO bands but did not detect them. The presence of ZrO bands would also be reflected in MS or S spectral types. However, the spectral type determinations of L$_2$~Pup available in the literature \citep{2014yCat....1.2023S} only list M-type classifications. We can, therefore, safely exclude that the companion to L$_2$~Pup is a white dwarf that shed material on the now observed red giant star. The interpretation that the companion is a K giant \citep{2015A&A...578A..77K} or a sub-stellar object \citep{2016A&A...596A..92K} is not excluded.

\section{Luminosity}\label{sec:Luminosity}

\begin{figure}[ht]
\includegraphics[width=\linewidth]{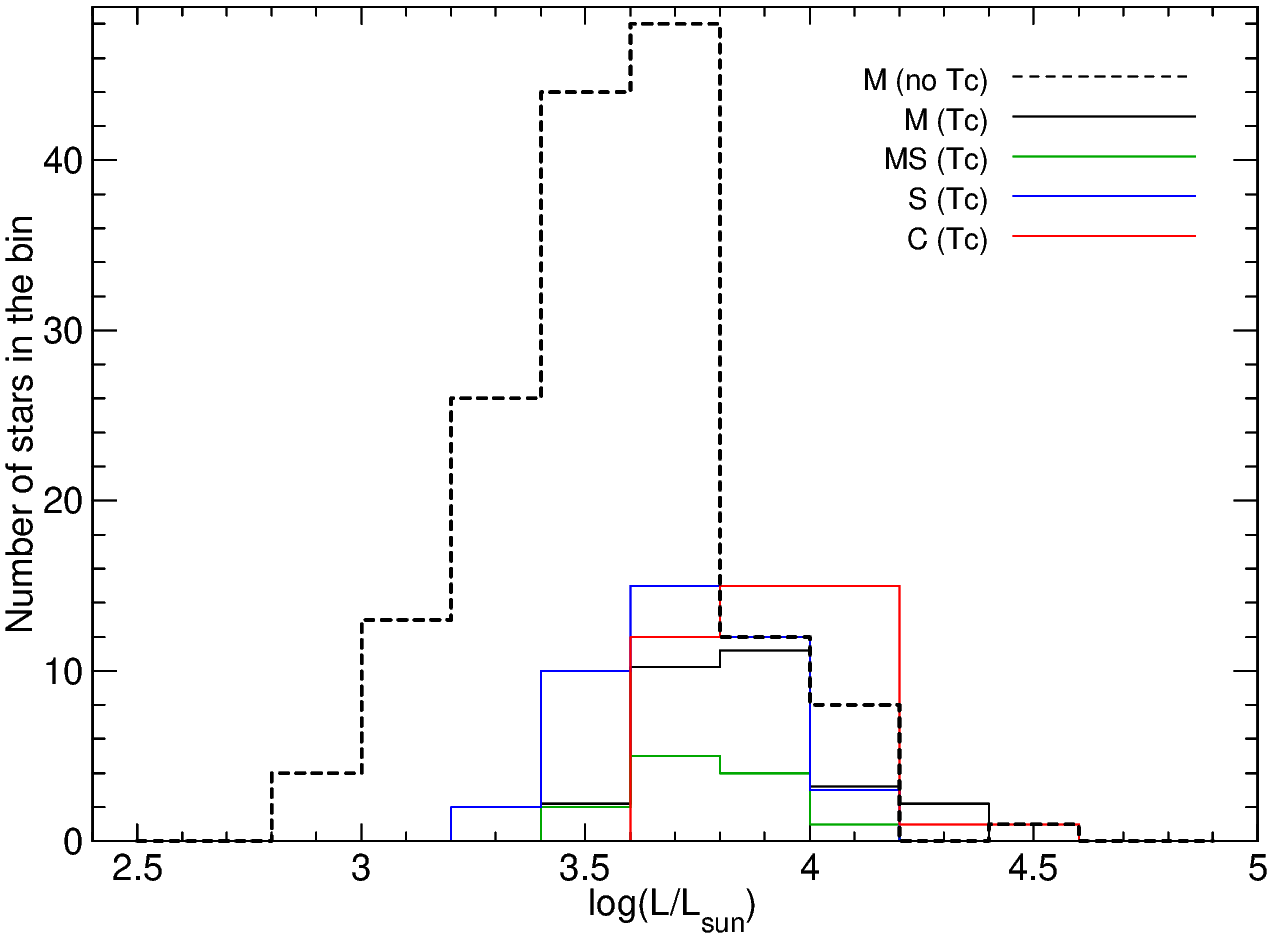}
\caption{Luminosity functions of the AGB comparison stars with Tc observations from \citet{2019A&A...622A.120U}.}
\label{fig:LF}
\end{figure}

The luminosity functions (LFs) of the comparison sample are shown in Fig.~\ref{fig:LF}. The Tc-poor M stars start at luminosities below 1000\,$L_{\sun}$ and increase in number with increasing luminosity, but this only reflects the selection function of the individual studies that observed these stars for Tc lines. However, the sudden decrease in their number around $\log(L/L_{\sun})=3.8$ is probably real and should be related to the luminosity when 3DUP has set in in most stars. It also roughly coincides with the peak of the Tc-rich sub-classes. The maximum luminosity is more interesting because these luminous, Tc-poor M-type giants could be candidates for intermediate-mass stars undergoing hot bottom burning. The absence of Tc is in agreement with the theoretical predictions that the $^{22}$Ne neutron source dominates in these stars \citep{2013A&A...555L...3G}. Some stars with $L>30\,000\,L_{\sun}$ exist in this group, which is actually higher than the most luminous C stars.

Although one would expect from evolutionary arguments stars to become brighter along the sequence M(Tc)--MS(Tc)--S(Tc), the mean luminosity actually decreases along that sequence. However, one must note that the number of stars is not large and it is probably fair to say that the mean luminosity does not increase along this sequence. Only the carbon stars are significantly brighter than the other Tc-rich groups, with a mean luminosity $\left<L\right>=9300\pm4350\,L_{\sun}$. This may be compared to the results of \citet{2022A&A...664A..45A}, who find $\left<L\right>=9180\pm4620\,L_{\sun}$ for their sample of 300 N-type carbon stars.

Tc-rich stars appear in significant numbers only at $L\geq2500\,L_{\sun}$. The faintest Tc-rich stars in the comparison sample are V915~Aql and Hen~4-19, for which we derive $L=1800\,L_{\sun}$ and $L=2500\,L_{\sun}$, respectively. Their low luminosities could be caused by a recent TP so that they are currently close to the post-TP luminosity minimum. V915~Aql has been confirmed to be Tc-rich by multiple literature sources \citep[e.g., ][]{1987AJ.....94..981L,1999A&A...351..533L,2019A&A...625L...1S}, hence there is no reason to doubt this classification. \citet{2019A&A...625L...1S} determined a luminosity of $L=1958\,L_{\sun}$ (based on its \Gaia DR2 parallax) and an initial mass close to $1\,M_{\sun}$. The same authors report additional Tc-rich stars at comparably low luminosities. However, the more precise \Gaia DR3 parallaxes place these stars at larger distances, so they must be more luminous than reported by \citet{2019A&A...625L...1S}, whereas V915~Aql's parallax remained almost unchanged in \Gaia DR3. Note also that V915~Aql must be a TP-AGB star because it is Tc-rich, although it does not fulfil the luminosity criterion of being brighter than the RGB tip. Hen~4-19 was determined to be Tc-rich by \citet{1999A&A...345..127V}. Observational constraints on the lower mass limit for 3DUP to occur are important for calibrating AGB evolutionary models because convective mixing is difficult to model from first principles. Stars such as V915~Aql and Hen~4-19 thus serve as important benchmark objects.

No parallax for L$_2$~Pup was available in the first and second \Gaia data releases. Earlier works that report its luminosity thus applied the \texttt{Hipparcos} parallax of $\varpi_{\rm HIP}=15.61\pm0.99$\,mas (equivalent to $d=64.1$\,pc) from the new reduction of \citet{2007A&A...474..653V}. The \Gaia parallax \citep[$\varpi_{G}=17.7906\pm0.8170$\,mas, equivalent to $d=56.2$\,pc,][]{2023A&A...674A...1G}, which we adopt here because of its higher precision, puts L$_2$~Pup closer to the sun, lowering its luminosity estimate. \citet{2022A&A...667A..74A} derive a larger distance of $100^{+14}_{-10}$\,pc, which would imply a much higher luminosity. However, they assign L$_2$~Pup to one of the OT modes and apply the PL relation of SRVs of \citet{2003A&A...403..993K}, which is probably not correct, see below. Based on the \Gaia DR3 parallax, we  derived a luminosity of $L=1490\pm150\,L_{\sun}$. We also collected luminosity estimates from the literature in Table~\ref{tab:L2Puplumi} and transformed them to the \Gaia-based distance by a factor $\left(\varpi_{\rm HIP}/\varpi_{G}\right)^2$. L$_2$~Pup could be as bright as the RGB tip luminosity only in the upper limit of the estimate by \citet{2020A&A...633A..34C} and adopting the \texttt{Hipparcos} parallax. However, their luminosity determinations are based on bolometric corrections in the $V$ and $I$ bands, which are uncertain for cool giant stars because they emit most flux in the IR. The weighted mean and its associated standard error are $\left<L\right>=1530\pm120\,L_{\sun}$.  We thus conclude that L$_2$~Pup is probably less luminous than the RGB tip. R~Doradus, the nearest AGB star, is also Tc-poor, but in our estimate has a luminosity of $\sim5600\,L_{\sun}$ clearly above the RGB tip, confirming its evolutionary status as an AGB star. 
\begin{table}
\caption{Luminosity determinations of L$_2$~Pup.}\label{tab:L2Puplumi}
\centering
\begin{tabular}{lrr}
\hline\hline
Reference & $L_{\rm HIP}$ &  $L_{\rm G}$ \\
          &($L_{\sun}$)   & ($L_{\sun}$) \\
\hline
K14       &  $2000\pm700$ & $1540\pm540$ \\
M17       &  $1845\pm314$ & $1420\pm240$ \\
C20       &  $2455\pm500$ & $1890\pm380$ \\
This work &               & $1490\pm150$ \\
\hline
Weighted mean &           & $1530\pm120$ \\
\hline
\end{tabular}
\tablefoot{$L_{\rm HIP}$: luminosity based on the \texttt{Hipparcos} parallax \citep{2007A&A...474..653V}; $L_{\rm G}$: luminosity and its uncertainty calculated with or transformed to the \Gaia DR3 parallax \citep{2023A&A...674A...1G}. References: K14: \citet{2014A&A...564A..88K}, M17: \citet{2017MNRAS.471..770M}, C20: \citet{2020A&A...633A..34C}.}
\end{table}

We can estimate the probability of a star at the approximate luminosity of L$_2$~Pup to be on the RGB or AGB from the rate of evolution $\Delta L/\Delta t$ from theoretical evolutionary tracks such as those of \citet{2022A&A...665A.126N}. We downloaded tracks of non-rotating PARSEC models with $Z=0.014$ and $Y=0.273$ from the authors' website\footnote{\url{http://stev.oapd.inaf.it/PARSEC/tracks_v2.html}} and calculated $\Delta L/\Delta t$ at luminosities between 1000 and 2000\,$L_{\sun}$. L$_2$~Pup's initial mass has been determined to be between $\sim1$ \citep{2016A&A...596A..92K} and $\sim2\,M_{\sun}$ \citep{2014A&A...564A..88K}. \citet{2020A&A...633A..34C} derived a mass of $1.54^{+1.44}_{-0.78}\,M_{\sun}$. We, therefore, inspected evolutionary tracks with initial masses between 1.0 and $1.5\,M_{\sun}$ and found that on the AGB\footnote{Note that the chosen luminosity range is covered by the horizontal branch tracks of the PARSEC models.}, $\Delta L/\Delta t$ is about 5.5 times larger than on the RGB. Stars on the AGB evolve faster than stars on the RGB by that factor, reducing the density of stars in the relevant luminosity range accordingly. This means that a randomly chosen, low-mass red giant star in the luminosity interval $1000-2000\,L_{\sun}$ has a probability of $\sim85$\% of being an RGB star and $\sim15$\% of being an AGB star.

%
%
%

\section{Pulsation properties}\label{sec:Pulsation} 

\begin{figure*}[ht]
\includegraphics[width=16cm]{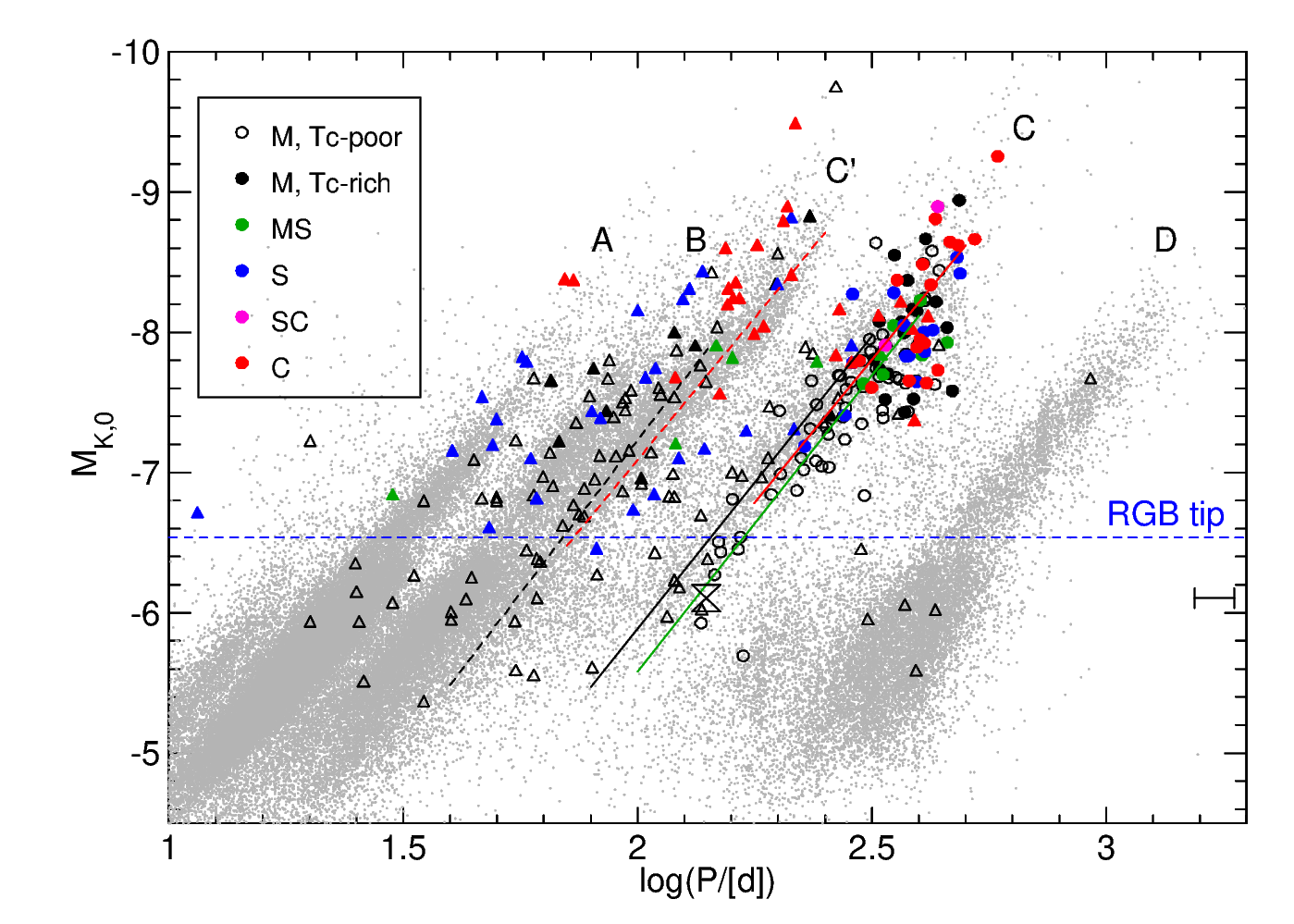}
\caption{Period -- absolute $K$ magnitude diagram of the comparison sample of \citet{2019A&A...622A.120U}. The symbol colours indicate the Tc classification and the chemical spectral type, see the legend. The symbol shape represents the variability type: circles for Miras and triangles for SRVs. The location of L$_2$~Pup is shown by the hourglass symbol. Over-plotted are LPV sequences from \citet{2007AcA....57..201S}: dashed lines: sequence C$^\prime$; solid lines: sequence C; black: O-rich stars; red: C-rich stars. The solid green line is the relation of Galactic O-rich Miras with $\log(P)<2.6$ found by \citet{2023MNRAS.523.2369S}. The dashed horizontal line indicates the RGB tip brightness ($M_{\rm K,0}=-6\fm54$). The horizontal bar on the right-hand side indicates the orbital period range of L$_2$~Pup's binary companion identified by \citet{2016A&A...596A..92K}, $P_{\rm orb}=4.69\pm0.45$\,yr. Grey dots in the background represent LPVs in the LMC from OGLE~III; only their primary periods and 2MASS magnitudes are plotted. A distance modulus of 18\fm477 to the LMC has been assumed to shift the LMC LPVs and the sequences to the absolute scale of the Galactic stars.}
\label{fig:PK0}
\end{figure*}

Figure~\ref{fig:PK0} shows the PL($K$) diagram of the comparison sample. The absolute $K$-band magnitude is used here as a proxy for the luminosity of the stars. The symbol colours indicate the Tc classification and the chemical spectral type; see the figure legend. The symbol shape, on the other hand, represents the variability type: circles indicate Miras, and triangles the SRVs. The RGB tip luminosity at $M_{\rm K,0}=-6\fm54$, indicated by the dashed horizontal line in Fig.~\ref{fig:PK0}, was calculated from the observed de-reddened RGB tip in the LMC \citep[$m_{\rm TRGB}(K_{\rm S})=11\fm94\pm0.04$,][]{2000A&A...359..601C} and the distance modulus to the LMC \citep[$\mu=18\fm477$,][]{2019Natur.567..200P}. Over-plotted are LPV sequences from \citet{2007AcA....57..201S}: dashed lines: sequence C$^\prime$; solid lines: sequence C; black: O-rich stars; red: C-rich stars. The same distance modulus to the LMC has been adopted to shift them to absolute $K$-band magnitudes. The solid green line in Fig.~\ref{fig:PK0} is the relation of Galactic O-rich Miras with $\log(P)<2.6$ found by \citet{2023MNRAS.523.2369S}.

PL relations of LPVs have been widely studied in the Magellanic Clouds \citep[MCs, e.g.,][]{2015MNRAS.448.3829W} but studies of Galactic stars are relatively rare because distances (parallaxes) of nearby AGB stars were available only in modest numbers and with significant uncertainties before the advent of \Gaia. As demonstrated by Fig.~\ref{fig:PK0}, the \Gaia DR3 parallaxes appear to be precise enough to reveal the PL sequences well-known from the MCs; see also \citet[][their Fig.~32]{2023A&A...674A..15L}. Many of the Galactic LPVs can be assigned unambiguously to one of the pulsation sequences, which are designated by the nomenclature of \citet{2015MNRAS.448.3829W} in Fig.~\ref{fig:PK0}. Typical uncertainties are generally much less than the width of the sequences themselves.

The Miras form the most obvious and best-separated sequence of all LPVs. They can all be assigned to the FM sequence~C. A tadpole structure seems to emerge: The majority of Miras forms the "head" at $\log(P)\sim2.5$, from which a "tail" extends to shorter periods and lower $M_{\rm K,0}$. The Miras in the "head" belong to all 3DUP stages, from no 3DUP to evolved C-rich Miras. Note that some of their periods are close to one year ($\log(P)\sim2.56$), their measured periods might be affected by seasonal gaps in their light curves in some cases. The Miras forming the "tail", on the other hand, are all Tc-poor, M-type Miras. They reside in a region where, at a given luminosity, typically lower-mass stars are located \citep[cf.][]{2021A&A...656A..66T}. 
Several SRVs can be assigned to the fundamental mode sequence, but most of them populate the OT mode sequences, in particular sequences~C$^\prime$ and B in the nomenclature of \citet{2015MNRAS.448.3829W}, both of which are attributed to pulsation in the first OT mode \citep{2017ApJ...847..139T}.

Sequence~D hosts candidates for long secondary periods (LSP). This phenomenon is suggested to be caused by binarity \citep{2021ApJ...911L..22S} or non-radial pulsations \citep{2002AJ....123.1002H}. For example, the $\sim2170$\,d period of $\alpha$~Ori has recently been suggested to be caused by a binary companion \citep{2024arXiv240809089G,2024arXiv240911332M}. Whereas essentially every pulsation sequence hosts all chemical spectral types, sequence~D is populated only by a few Tc-poor SRVs of the comparison sample. Only one of them is brighter than the RGB tip. A larger sample of LSP candidates would need to be observed to check if they can be TP-AGB stars undergoing 3DUP.

Tc-poor stars can be found at any luminosity, even at the same location as the carbon stars. Their number drops sharply at $M_{\rm K,0}\approx-8\fm0$, but they do not disappear even at the highest luminosities. Possibly, Tc-poor stars brighter than this threshold are intermediate-mass AGB stars undergoing HBB (see Sect.~\ref{sec:Luminosity}).

The number of S stars pulsating in the OT modes may appear large compared to those on the fundamental mode sequence, but this is most likely a selection bias because several of the literature sources from which Tc observations were collected avoided Miras on purpose to facilitate abundance determinations \citep[e.g.,][and references therein]{2021A&A...650A.118S}. The faintest S star is again V915~Aql at $M_{\rm K,0}=-6\fm46$. It is the only Tc-rich star fainter than the RGB tip luminosity. Several faint S stars are only marginally brighter than the RGB tip.

Irrespective of pulsation mode, all carbon stars are much brighter than the faintest S stars. They are the most luminous group (cf.\ Table~\ref{tab:lumi}) and appear in significant numbers only at $M_{\rm K,0}\lesssim-7\fm5$. Their mean absolute magnitude is $\left<M_{\rm K,0}\right>=-8.22\pm0.47$\,mag. This may be compared again to the results of \citet{2022A&A...664A..45A}, who find $\left<M_{\rm K,0}\right>=-8.16\pm0.57$\,mag for their sample of 300 N-type carbon stars.

An interesting detail of Fig.~\ref{fig:PK0} is that the faintest Tc-rich stars are SRVs, whereas Tc-rich Miras appear only at $M_{\rm K,0}\lesssim-7\fm4$; the lower part of the fundamental mode sequence~C is occupied by Tc-poor Miras and SRVs. This may be due to two effects. i) The more massive stars ($M>2\,M_{\sun}$) might undergo 3DUP when they are still pulsating in one of the OT modes \citep{1999A&A...351..533L}. However, it is not clear if this would happen at such low luminosities. ii) Stars could switch from the fundamental to the OT mode when they experience a TP and 3DUP event, and we might happen to observe them during the luminosity minimum after the onset of a TP. This was speculated to be the case for V915~Aql, cf.\ Sect.~\ref{sec:Luminosity}. However, this would be expected to affect only about 1\% of Miras at any one time \citep{1981ApJ...247..247W}, and so these stars would be significantly over-represented in the comparison sample. Some of the comparison sample stars were collected from studies that specifically targeted stars with changing pulsation periods that could result from a recent TP, e.g., \citet{2011A&A...531A..88U}. However, these authors studied Mira stars, not SRVs.

L$_2$~Pup is represented in Fig.~\ref{fig:PK0} by the hourglass symbol, reminiscent of its bipolar circumstellar structure. Its well-defined pulsation period of $\sim140$\,d has been remarkably stable in the interval 135 to 145\,d for almost a century, unaffected by the 1995 dimming event \citep{2002MNRAS.337...79B}. L$_2$~Pup is classified as an SRa variable with a full amplitude of $\Delta V\sim1\fm2$. This variable class is thought to be similar to Mira variables, only with amplitudes $\Delta V<2\fm5$. \citet{2020MNRAS.499.4687C} characterised the power spectral density properties of L$_2$~Pup and concluded that it is a coherent (classic) pulsator driven by coherent perturbations, similar to the Mira variable U~Per, as opposed to a stochastic (or solar-like) pulsator. It has been discussed by \citet{2009A&A...498..489J} and \citet{2014A&A...564A..88K} if the photometric 140\,d period could be caused by a close-in binary companion, but this seems unlikely. The binary companion identified by \citet{2016A&A...596A..92K} has an orbital period of $P_{\rm orb}=4.69\pm0.45$\,yr, which is clearly outside the range of Sequence~D at L$_2$~Pup's luminosity; this is indicated by the horizontal bar in Fig.~\ref{fig:PK0}.

As pointed out in the Introduction, we adopted the COBE/DIRBE observations from \citet{2004ApJS..154..673S} in the $2.2\,\mu{\rm m}$ band to calculate L$_2$~Pup's pre-dimming absolute magnitude of $M_{\rm K,0}=-6\fm17$. Thus, L$_2$~Pup was fainter than the RGB tip (dashed blue line) already in the pre-dimming phase. On the other hand, it exactly coincides with the relation of Galactic O-rich Miras derived by \citet[][solid green line in Fig.~\ref{fig:PK0}]{2023MNRAS.523.2369S}. As pointed out by \citet{2023MNRAS.523.2369S}, the local O-rich Miras are fainter than the LMC counterparts, cf.\ the solid black line in Fig.~\ref{fig:PK0} for the relation of sequence~C from \citet{2007AcA....57..201S}. L$_2$~Pup's position in the diagram is fully consistent with it being an FM pulsator, and several other Galactic Tc-poor Miras and SRVs are located in its vicinity. However, it is clearly disconnected from the Tc-rich variables further up on sequence~C.

Also included as grey dots in the background of Fig.~\ref{fig:PK0} are LPVs in the LMC from the OGLE~III catalogue \citep{2009AcA....59..239S}. Their number density drops noticeably above the RGB tip. \citet{2004AcA....54..129S} showed that small-amplitude variable red giants below the RGB tip in the MCs are a mixture of RGB and AGB stars. They also devise a method to separate stars on the two giant branches, but this method works only for stars pulsating in higher-order modes. Separating RGB and AGB stars on the FM sequence~C is not possible, and the nature of the few FM pulsators below the RGB tip is unclear. We note that in Fig.~\ref{fig:PK0}, this sequence is poorly populated below the RGB tip compared to the higher-order modes and sequence~D.

Unfortunately, L$_2$~Pup's pulsation properties do not give us clear additional clues to its evolutionary state. However, its similarity to short-period, low-luminosity Miras favours a low initial mass, in agreement with its membership in the thick disc \citep{2002ApJ...569..964J}. This aligns well with the slight metal deficiency of $[{\rm Fe}/{\rm H}]=-0.4$ found by \citet{2020A&A...633A..34C}.
\section{\Gaia-2MASS diagram}\label{sec:Gaia2MASS}

\begin{figure}[ht]
\includegraphics[width=\linewidth]{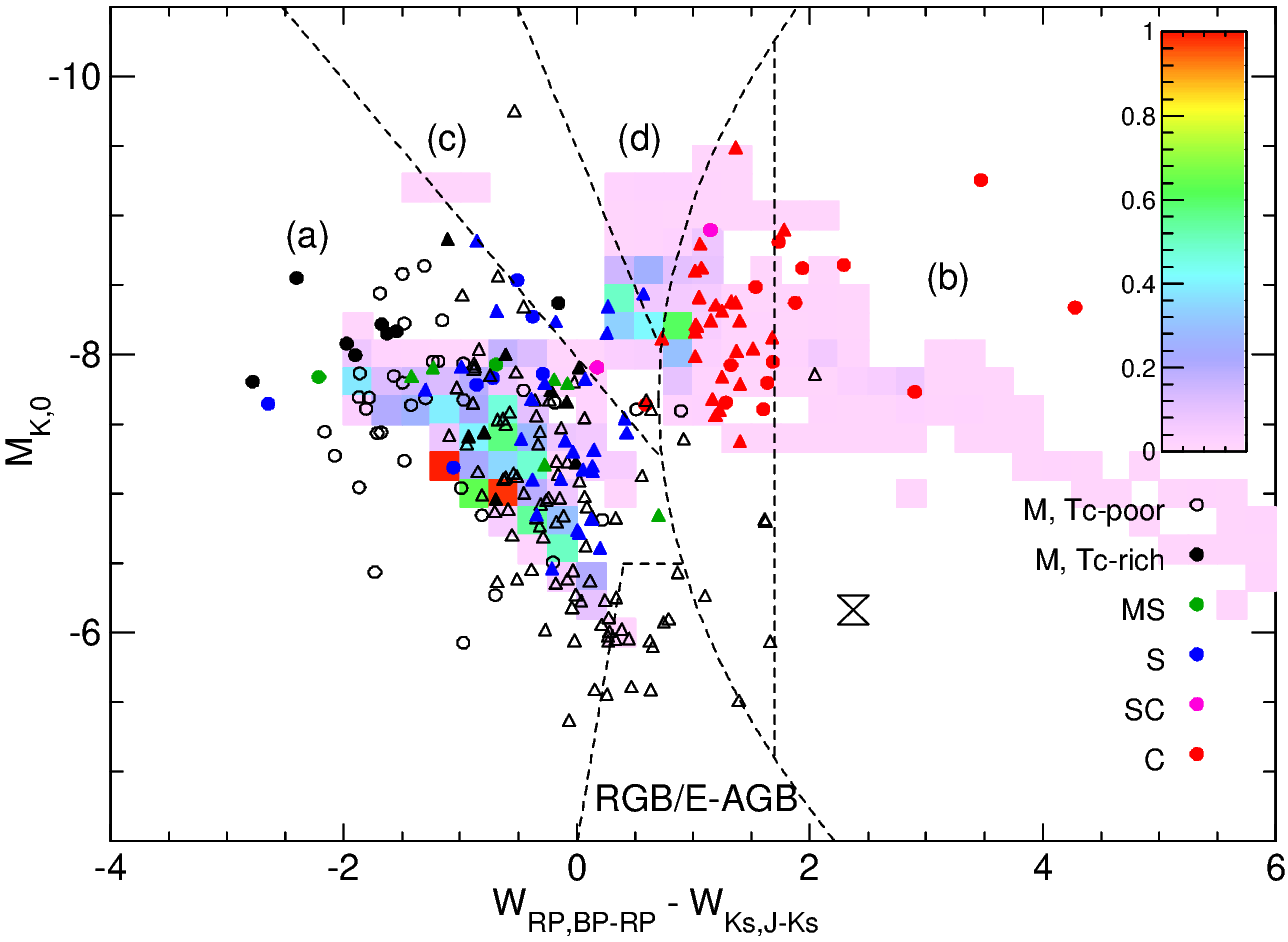}
\caption{\Gaia-2MASS diagram of Galactic LPVs, including L$_2$~Pup. Symbols as in Figure~\ref{fig:PK0}. The colour bar on the right-hand side indicates the relative density of AGB evolutionary model grids calculated with the \texttt{COLIBRI} code \citep{2013MNRAS.434..488M}.}
\label{fig:G2M}
\end{figure}

Finally, we inspect L$_2$~Pup and the AGB comparison sample in the \Gaia-2MASS diagram in Fig.~\ref{fig:G2M}. The \Gaia-2MASS diagram is a relatively new method to distinguish various types and masses along the AGB. It plots the (absolute) magnitude in the $K_{\rm S}$ band as a function of the difference between two Wesenheit indices. Wesenheit indices are reddening-free quantities defined as a linear combination of the magnitude and colour of a star. The \wbr and \wjk indices have been defined by \citet{2018A&A...616L..13L} and \citet{2005AcA....55..331S}, respectively, and are obtained from the \Gaia and NIR photometry in the following ways:
\begin{equation}
\wbr = G_{\rm RP} - 1.3\cdot(\bpminrp)
\end{equation}
\begin{equation}
\wjk = K_{\rm S} - 0.686\cdot(J - K_{\rm S}).
\end{equation}
\begin{equation}
\wkr = \wbr-\wjk
\end{equation}

We follow the definition of \citet{2018A&A...616L..13L} for naming the different regions in this diagram:

\begin{itemize}
    \item RGB and early (faint) AGB: This is the region at \wkr between about 0 and 2, in the lower part of the diagram;
    \item (a): Low-mass, oxygen-rich AGB stars;
    \item (b): C-rich AGB stars, where those to the right of the vertical line at $\wkr=1.7$ are extreme C stars;
    \item (c): Intermediate-mass AGB stars;
    \item (d): Massive AGB ($M_{i}\gtrsim5\,M_{\sun}$) and RSG ($M_{i}\gtrsim8\,M_{\sun}$) stars.
\end{itemize}

The density map in the background of Fig.~\ref{fig:G2M} represents an AGB evolutionary model grid calculated with the \texttt{COLIBRI} code of \citet{2013MNRAS.434..488M} with metallicity $[{\rm Fe}/{\rm H}]=-0.03$ and masses 1.0, 1.3, 1.5, 1.8, 2.0, 2.4, 2.6, 3.0, 4.0, and 5.0\,$M_{\sun}$. The $M_{\rm K, 0}$ vs.\ \wkr plane was divided into bins of 0\fm20 width in $M_{\rm K, 0}$ and 0\fm25 in \wkr. For each bin, the time a model spent in that bin was counted to represent the probability of finding a model star of the given mass in that stage. The evolutionary tracks are weighted by their mass according to a Kroupa IMF with slope $\alpha=-2.3$ \citep{2001MNRAS.322..231K}, assuming a constant star formation rate. In addition, model grid points have been selected such that the stars have evolved to or past the point at which the first OT mode has become dominant \citep[see][for a description of how stars progress through the various pulsation modes and sequences]{2017ApJ...847..139T}. For OT modes, growth rates are a good proxy for variability amplitudes \citep{2017ApJ...847..139T}. To a first approximation, the variability amplitudes depend on the ratio between the dynamical frequency and the acoustic cut-off frequency of the star, $\tilde{\nu}$. According to the definition in Appendix~B.2 of \citet{2018A&A...616L..13L}, we adopt the selection criterion $\tilde{\nu}=(R T_{\rm eff}/M)^{1/2}>10$, with all stellar parameters in solar units. The density of the grid was scaled to a maximum of 1 in Fig.~\ref{fig:G2M}.

The model grid well reproduces the locations of the observed sample stars, in particular in the branch (a) (low-mass, O-rich AGB stars). Most of the sample stars, even those with 3DUP, appear to be low-mass stars on branches (a) and (b). This is in agreement with recent mass estimates of AGB stars from the oxygen isotopic ratios, see e.g.\ \citet{2017A&A...600A..71D}, and indeed expected from the IMF. Only a few stars are found in branch (c) of intermediate-mass AGB and in branch (d) of massive AGB and RSG stars. Most of the stars in branch (c) are close to the border to branch (a); they might be placed in branch (c) only because of their variability or photometric uncertainty.

The observed carbon stars are all in branch (b), confirming the suitability of the \Gaia-2MASS diagram to distinguish chemical types. The extreme C-star branch, to the right of the region (b), is sparsely populated by the AGB comparison sample, which can be well explained by the way the sample was selected from optically non-obscured stars because the Tc lines in the blue spectral region need to be observable. Essentially all stars in that region of branch (b) are Miras. This agrees with the results of \citet{2019gaia.confE..62M}, who find that LMC stars with such high \wkr have large amplitudes and are thus Mira candidates. There are two SC-type stars in the sample, R~CMi and RZ~Peg. The former is located at the border between the O-rich and the C-rich regime, and the latter in branch (c), at larger \wjk than most S stars. This is in agreement with the results of \citet{2022A&A...664A..45A} and is also what would be expected from the evolutionary state of SC stars intermediate between S and C stars.

L$_2$~Pup is represented by the hourglass symbol in Fig.~\ref{fig:G2M}. As mentioned in the Introduction, we adopted the post-dimming near-IR photometry of \citet{2002MNRAS.337...79B} to calculate the \wjk index to be more consistent with the \Gaia photometry. The $M_{\rm K,0}$ magnitude, on the other hand, reflects the pre-dimming brightness observed by the COBE/DIRBE instrument. Again, L$_2$~Pup's position in the \Gaia-2MASS diagram is consistent with being less luminous than the RGB tip, indicated by the upper limit of the RGB and early (faint) AGB branch in the \Gaia-2MASS diagram. However, its \wkr index is much larger than that of the other Tc-poor SRVs in the comparison sample at comparable brightness, shifting it to the branch (c) of the C-rich AGB stars. However, it is not the only Tc-poor, M-type SRV in that region of the diagram, only the one with the largest \wkr index of that group. As suggested by \citet{2002MNRAS.337...79B}, significant extinction existed even before the 1995 dimming event. This circumstellar extinction probably is also the cause of L$_2$~Pup's large \wkr index. The chemical model of the disc developed by \citet{2024MNRAS.532..734V} predicts carbon chemistry arising from the disc structure and therefore some carbonaceous dust may be formed, which could contribute to the circumstellar reddening. Other Tc-poor, M-type SRVs with large \wkr indices could be candidates for having circumstellar discs similar to that round L$_2$~Pup.

\section{Conclusions}\label{sec:Conclusio}

We revised the evolutionary state of the nearby red giant star L$_2$~Puppis in comparison with a sample of known Galactic AGB stars. Most importantly, we presented a new analysis of archive spectra of L$_2$~Pup for the third dredge-up indicator technetium (Sect.~\ref{sec:Tc}), demonstrating that it is Tc-poor and lifting the ambiguity in the literature. We calculated luminosities of L$_2$~Pup and the comparison sample stars (Sect.~\ref{sec:Luminosity}), finding that it is fainter than all known Tc-rich TP-AGB stars in the comparison sample and fainter than the RGB tip luminosity. The analysis of the pulsation properties (Sect.~\ref{sec:Pulsation}) confirms earlier conclusions that L$_2$~Pup pulsates in the fundamental radial mode. However, the pulsation properties do not allow us to clearly distinguish between an RGB or AGB evolutionary state. The \Gaia-2MASS diagram confirms that L$_2$~Pup had significant circumstellar extinction before the dimming event.

In conclusion, L$_2$~Pup could be an RGB or early AGB star. It cannot be a TP-AGB star because it lacks Tc in its atmosphere and is fainter than the RGB tip. Based on purely statistical arguments from the evolutionary rate $\Delta L/\Delta t$ at L$_2$~Pup's luminosity of $L\approx1530\,L_{\sun}$, the probability that it is an AGB star (15\%) is smaller than that for being an RGB star (85\%). In any case, it is not safe to call L$_2$~Pup an AGB star, as is often done in the scientific and non-scientific literature. Considerations of its evolutionary state have consequences on our understanding of the evolution of the whole system including a circumstellar disc and a companion.

\begin{acknowledgements}
This paper resulted from discussions at the Advances in Cool Evolved Stars (ACES) Conference at Monash University, Melbourne, Australia, from July 08 to 12, 2024. The author would like to thank the conference organisers and the debaters, in particular T.\ Danilovich for detailed comments on the manuscript. This research was funded in part by the Austrian Science Fund (FWF) 10.55776/F81. For open access purposes, the author has applied a CC BY public copyright license to any author accepted manuscript version arising from this submission. We acknowledge with thanks the variable star observations from the AAVSO International Database contributed by observers worldwide and used in this research. This work has made use of data from the European Space Agency (ESA) mission \Gaia (\url{https://www.cosmos.esa.int/gaia}), processed by the \Gaia Data Processing and Analysis Consortium (DPAC, \url{https://www.cosmos.esa.int/web/gaia/dpac/consortium}). Funding for the DPAC has been provided by national institutions, in particular the institutions participating in the \Gaia Multilateral Agreement. This publication makes use of data products from the Two Micron All Sky Survey, which is a joint project of the University of Massachusetts and the Infrared Processing and Analysis Center/California Institute of Technology, funded by the National Aeronautics and Space Administration and the National Science Foundation. This publication makes use of data products from the Wide-field Infrared Survey Explorer, which is a joint project of the University of California, Los Angeles, and the Jet Propulsion Laboratory/California Institute of Technology, funded by the National Aeronautics and Space Administration.
\end{acknowledgements}

\bibliographystyle{aa}
\bibliography{References}

\end{document}